\documentclass[pdflatex,sn-chicago]{sn-jnl}

\usepackage{graphicx}%
\usepackage{multirow}%
\usepackage{amsmath,amssymb,amsfonts}%
\usepackage{amsthm}%
\usepackage{mathrsfs}%
\usepackage[title]{appendix}%
\usepackage{xcolor}%
\usepackage{textcomp}%
\usepackage{manyfoot}%
\usepackage{booktabs}%
\usepackage{algorithm}%
\usepackage{algorithmicx}%
\usepackage{algpseudocode}%
\usepackage{listings}%
\usepackage{multirow}

\theoremstyle{thmstyleone}%
\theoremstyle{thmstyletwo}%

\theoremstyle{thmstylethree}%

\graphicspath{{figures/}}

\begin{document}

\title[Conceptual Review]{Human-Centered Design for AI-based Automatically Generated Assessment Reports: A Systematic Review}


\author[1]{\fnm{Ehsan} \sur{Latif}}\email{ehsan.latif@uga.edu}
\author[2]{\fnm{Ying} \sur{Chen}}\email{ychen406@uillinois.edu}
\author*[1]{\fnm{Xiaoming} \sur{Zhai}}\email{xiaoming.zhai@uga.edu}
\author[3]{\fnm{Yue} \sur{Yin}}\email{yueyin@uic.edu}


\affil*[1]{\orgdiv{AI4STEM Education Center}, \orgname{University of Georgia}, \city{Athens}, \postcode{30602}, \state{GA}, \country{USA}}

\affil[2]{\orgdiv{Illinois Workforce Educational Research Collaborative}, \orgname{University of Illinois System}, \city{Chicago}, \postcode{60606}, \state{IL}, \country{USA}}

\affil[3]{\orgdiv{Department of Educational Psychology}, \orgname{University of Illinois Chicago}, \city{Chicago}, \postcode{60607}, \state{IL}, \country{USA}}


\pretocmd{\abstractname}{\newpage}{}{}

\abstract{This paper provides a comprehensive review of the design and implementation of automatically generated assessment reports (AutoRs) for formative use in K-12 Science, Technology, Engineering, and Mathematics (STEM) classrooms. With the increasing adoption of technology-enhanced assessments, there is a critical need for human-computer interactive tools that efficiently support the interpretation and application of assessment data by teachers. AutoRs are designed to provide synthesized, interpretable, and actionable insights into students' performance, learning progress, and areas for improvement. Guided by cognitive load theory, this study emphasizes the importance of reducing teachers' cognitive demands through user-centered and intuitive designs. It highlights the potential of diverse information presentation formats such as text, visual aids, and plots and advanced functionalities such as live and interactive features to enhance usability. However, the findings also reveal that many existing AutoRs fail to fully utilize these approaches, leading to high initial cognitive demands and limited engagement. This paper proposes a conceptual framework to inform the design, implementation, and evaluation of AutoRs, balancing the trade-offs between usability and functionality. The framework aims to address challenges in engaging teachers with technology-enhanced assessment results, facilitating data-driven decision-making, and providing personalized feedback to improve the teaching and learning process.}




\maketitle

\section{Introduction}\label{sec:introduction}

Classroom assessments increasingly employ sophisticated technologies, such as artificial intelligence (AI), simulations, games, videos, and visualizations to advance teachers’ instructional decision-making and engage students in meaningful learning. Utilizing the results of technology-enhanced assessments, educators can gauge learning processes and assess complex constructs that traditional classrooms cannot offer \citep{liu2016validation,zhai2023technology,guo2024using}. While teachers are provided with growing information to support instruction, they tend to be overloaded with this information. The more does not always indicate the better \citep{lee2023gemini}. Useful information must be sharp, straightforward, and cognitive-relief for teachers's instructional decision-making. In addition, with these sophisticated technologies and assessment outcomes \citep{latif2024fine}, most teachers have insufficient pedagogical content knowledge to interpret and transform the assessment outcomes into instructional practices \citep{joram2020influences,lee2024applying}. 

To support both teachers and students in using the classroom assessment results, researchers highlight the importance of the design principles for computer-human interfaces to optimize the amount, depth, and format of assessment results presented to users \citep{wu2024unveiling}. \cite{latif2024autoR} defined the design principles for instruction systems considering the teacher as a user of the system, including visualization, goal orientation,  information analysis, and personalized learning. The digital reporting systems can provide teachers with students’ performance and various learning metrics, including automatic scoring, longitudinal learning progress, identification of learning gaps, and personalized feedback \citep{carroll2021their,latif2024automatic,latif2024physicsassistant}. This paper defines this digital interface as an automatic assessment report (AutoR). The AutoR serves as a centralized hub for efficient and automatic analysis of classroom assessment data \citep{lamar2013using}, as well as data-driven decision-making tools for teachers\citep{aljohani2013learning, mottus2015use,latif2024can}.

To unlock the full potential of AutoRs, it is essential to deliver interpretable, synthesized, and sufficiently informative results for teachers. Previous studies have raised concerns about the design of an effective AutoR, as \cite{kasepalu2022teachers} pointed out that presenting teachers with unprocessed and unfamiliar information would diminish their engagement with the AutoR. Similarly, \cite{echeverria2018exploratory} suggested the importance of a narrative-driven approach in presenting classroom results, ensuring that the visualization aligns coherently with the learning goals. Although these findings are enlightening for the AutoR design, a comprehensive conceptual framework is needed \citep{sedrakyan2019guiding}. Such a framework would not only guide the design of AutoRs but also provide a structured approach for their implementation.

This study aims to develop a human-centered framework that conceptualizes the design and implementation of AutoRs, with a specific emphasis on their applications in K-12 Science, Technology, Engineering, and Mathematics (STEM) education. Using the conceptual framework, we conduct an in-depth review of the design elements of the most recent AutoRs adopted in K-12 STEM classrooms. These AutoRs consist of digital platforms that are publically available for analysis and the required features of AutoR based on our designed framework. This review study aims to address the following research questions: 
\begin{enumerate}
    \item What are the design features of existing AutoRs for K-12 STEM education in terms of cognitive demands and human-centered design?
    \item How does the design of AutoRs address the cognitive demands of teachers in using the AutoRs?
    \item How is AI incorporated into the AutoR and enhance the effectiveness?

\section{AutoR Design}\label{sec:autoR}

AutoR is a subset of the dashboard that offers an interface for users to interact with digital equipment. AutoR is uniquely positioned in educational assessment practices to support learning and instruction. AutoRs can display various types of data, such as raw student responses, demographic information, statistical data, and log data of learners’ interaction with the learning system \citep{latif2024autoR}. The diverse data sources may be analyzed using different methods and strategies before being displayed on AutoRs. For example, Rasch models and machine learning can be used to analyze classroom assessment data \citep{zhai2022applying}. Analyzed data with the information encapsulated are usually synthesized and presented on AutoRs for specific purposes.

Design of AutoRs is an evolving and ongoing process informed by the purpose of uses, users, data sources, and data visualization techniques \citep{garcia2020learning}. In classroom settings, AutoRs are frequently used to support teachers' instructional decision-making or provide students with customized learning feedback or support \citep{gerard2022supporting}. For the prior uses, the information provided in AutoRs should be straightforward and easily interpreted, given teachers' limited time to interpret the information and translate it into practice \citep{xu2020applying}. For the latter, AutoRs should be designed considering learners' performance levels, individual differences, and cultural backgrounds \citep{shemshack2020systematic}. For example, for students with learning disabilities, AutoRs may need to incorporate reading assistance \citep{Panjwani2023ai}. Despite the powerful functionality of AutoRs, the design of AutoRs should avoid an overwhelming amount of data provided to users, considering humans' working memory limitations \citep{yigitbasioglu2012review,latif2023ai}. Further, to support users to easily and accurately interpret the assessment results and utilize the assessment results to promote teaching and learning, visualizations of the data are critical in reducing teachers' cognitive loads \citep{Sedrakyan2019-zv,michaeli2020teachers}. 
 
Researchers have studied and synthesized how to design easy, accurate, and helpful AutoRs. \cite{duval2011attention} traced user attention to explain how the visualization of data and the recommendation function can be used to increase awareness and provide support. \cite{verbert2014learning} highlighted the purpose of supporting awareness, reflection, sense-making, and impact on learning of AutoR design. \cite{schwendimann2016perceiving} reviewed learning dashboard research and concluded that the dashboard lacks a design that differentiates it from other areas and addresses the activities of learning and teaching. Researchers have considered cognition theories such as working memory, situational awareness, motivation, and sensemaking as guiding principles in dashboard development \citep{yoo2015educational,verbert2020learning,echeverria2018exploratory}. These studies have established the foundation of cognition theories in orienting the design and presentation of the data and highlighted multiple considerations to guide the design of the AutoR. It is, however, necessary to consolidate these considerations into a conceptual framework and evaluate the design of existing AutoR, given the challenges of real-time usage. 

\section{Human-Centered Design for AutoR}
It is challenging for teachers to process the real-time information generated in an assessment system in a timely fashion. Teachers’ working memory limits their capacity to interpret AutoRs \citep{yoo2015educational}, particularly when the information provided by AutoR is rich. Therefore, it is essential to develop visualizations that are interpretable immediately with clear considerations of the users and the particular purposes of uses \citep{duval2011attention}. In practice, thoughtful design should consider users’ cognitive load and promote teachers to effectively use AutoRs \citep{kasepalu2022teachers}. The challenge of interpreting AutoRs in a short time motivates us to incorporate cognitive load theory in the framework for AutoR design. 

Cognitive load theory was first introduced as an instructional design theory based on human cognitive architecture and has to be considered for the design of educational technologies \citep{sweller2019cognitive,sweller2020cognitive}. This theory has been used broadly in human-computer interaction research since cognitive processing is affected by the content, presentation, and interactivity \citep{hollender2010integrating}. There are three categories of cognitive load: intrinsic, extraneous, and germane. \citet{sweller2019cognitive} suggested that a) intrinsic cognitive load is decided by the complexity of information processing and the person's ability to process the information; b) the extraneous cognitive load is influenced by the presentation of the information and the procedure that a person needs to interact with the presentation; and c) the germane cognitive load is intertwined with intrinsic cognitive load, which refers to the necessary working memory resources to learn. 

Following the categories of cognitive load, the goal of designing AutoR is to reduce the extraneous cognitive load by controlling the complexity of the information \citep{skulmowski2022understanding}. In cognitive load theory, informational complexity is measured by element interactivity \citep{sweller2020cognitive}. Element interactivity can be further divided into the number of elements and the interacting relationships between the elements in processing the information. In the design of AutoRs, we refer to the element interactivity to (a) the information that teachers receive simultaneously while interpreting the AutoR, (b) how the information is synthesized to make instructional meanings, and (c) how AI is embedded in AutoR for automatic reporting. Following these arguments, we proposed a conceptual framework for teacher-centered AutoRs design, including cognitive demands, AI embodiment and human-centered design support (see Fig.~\ref{fig:design_framework}). 

\begin{figure}
    \centering
    \includegraphics[width=1\linewidth]{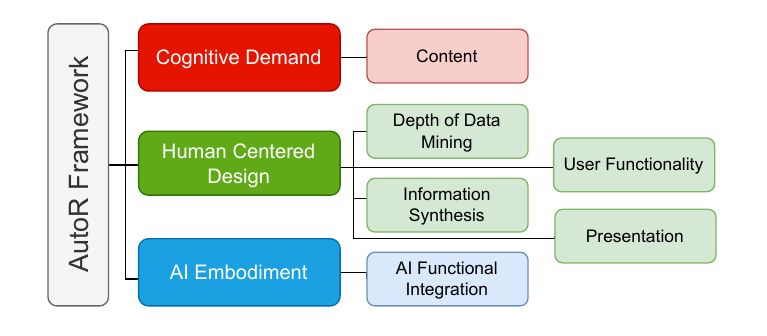}
    \caption{The design framework of AutoRs consisting three dimentions (Cognitive demand, AI embodiment, and Human centered design).}
    \label{fig:design_framework}
\end{figure}

\subsection{Cognitive Demands}
The dimension of cognitive demands examines the amount and level of integration of the information in the report. Three characteristics are recognized to count the amount and the level of the information: 1) the content presented in the report, 2) the synthesis level of presented information, and 3) the depth of data mining.

    \textbf{Content.} The intended content presented in the AutoR is expected to be the major driver of the cognitive load. Generated by computer, there is a lot of data, such as students' personal information, actions, and time tracking, that have educational significance \citep{nyland2018review}. Research has also found rich information that is helpful for teachers to know, such as learning progressions, scientific practices performance, standards matching, etc. \citet{zhai2021practices} pointed out that it is meaningful for teachers’ assessment practices to have the information that refers to the collecting and concluding evidence of students' current learning and providing feedback for future decisions and actions. In the characteristic of content, we detect what type of information is considered meaningful by the designers to present to teachers.
    
   \textbf{Information Synthesis Level.} Information synthesis indicates the width of the information. Although the width of the information can increase the cognitive load for users, sometimes they are necessary. \citet{sedrakyan2019guiding} suggest that teachers often need to compare students’ performance to identify critical information. For example, presenting information about a group usually indicates a higher level of synthesis than that of an individual, as information about a group is much richer than for the individual. However, such rich information can be helpful for teachers as it automatically undertakes part of the work that teachers will do when facing individual information. For example, teachers may organize different learning activities for different groups.
   
    \textbf{Depth of Data Mining.} Similar to the information synthesis level, the depth of data mining also describes how information included in the report is generated. Data mining with depth, although increasing cognitive load, might be helpful for teaching and learning. For example, student ability measures calculated from the Rasch model are deeper and more sophisticated than raw scores, but they can convey information about students’ growth over time and allow comparison of students taking different tests. Both the synthesis level and depth of data mining could alter the cognitive needs of users.

\subsection{Design Support}
The dimension of design support examines the design features that can assist teachers in interpreting and using the information. We identified the following two sub-dimensions:

\textbf{User Functionality:} User functionality refers to the capabilities and features that are provided for additional uses in class besides data analysis and processing. For formative assessment, it is widely accepted that timely and informative feedback is critical and helpful for learning \citep{narciss2008feedback}. Such feedback not only provides relevant data results (scores, grades, etc.) but also facilitates learning engagement, efficiency, and effectiveness with personalized and constructive information \citep{spector2016technology}. Correspondingly, the functionality of the AutoRs can expand from providing data results to support learners’ individualized learning engagement, for example, the function design of interactions between teacher and students \citep{spector2008editorial}. To provide information with efficiency, a common example function used in AutoRs is to use filters or tags to help teachers swiftly find the information needed.

\textbf{Information Presentations:} Information presentation refers to the forms of data visualization that help teachers interpret the report. As a branch of data analysis and processing, data visualization converts abstract and complex analytic results to concrete and visible information by amplifying human cognition \citep{yoo2015educational,card1999readings}. Common visual elements include charts, graphs, indicators, and alert mechanisms \citep{podgorelec2011taking,yoo2015educational}. Each element can reduce users' cognitive load in some ways. For example, charts are used to expose relationships between factors and indicators are used to highlight important information,\citep{barana2019learning, nyland2018review}.

It is unclear how the current AutoRs are designed and whether and how the design of AutoRs addresses the cognitive demand of each AutoR. To fill in this gap, this study reviews existing AutoRs using AutoR's design framework. The information can help evaluate the current AutoR and guide the future AutoR design.

\end{enumerate}
\subsection{AI intergration}
This dimension reveals a multifaceted approach to integrating AI in educational settings. AI technologies are employed to automate grading and provide performance analytics, thus reducing teachers’ administrative workload and allowing them to focus on more impactful instructional activities. The AI embodiment and functions, as described by developers, emphasize minimizing extraneous load through streamlined interfaces and clear, concise feedback mechanisms. By leveraging AI to manage cognitive load effectively, these systems enhance the learning experience for students and support teachers in delivering more efficient and effective instruction.

\begin{itemize}
    \item\textbf{AI embodiment/AI function} AI embodiment often includes user-friendly interfaces and intuitive designs that minimize unnecessary information and distractions. The embodiment can also manage teachers' intrinsic load by breaking down complex information into simpler, more digestible components. In this aspect, we analyze the developer's descriptions of whether the AutoRs have the function of streamlined interfaces, automating routine tasks, breaking down complex information, offering personalized feedback, etc.
    \item\textbf{AI integration} The integration of AI in AutoR is designed to optimize instructional(Predictive analytics reduce the extraneous load by proactively addressing potential learning difficulties) and administrative tasks (significantly reduces the extraneous load for teachers by eliminating repetitive, time-consuming tasks), as described by developers. These tasks are strategically aligned with the Cognitive Load Theory to ensure efficient use of cognitive resources. We examine whether the AI in AutoR serves as a standalone function or works organically with the AutoR system as the developers described. 
    \item\textbf{AI function} A binary codes that differentiate the function of AI in AutoR according to the developers' intention: AI functionalities directly impacting students, such as personalized feedback and adaptive learning paths; or AI functionalities directly impacting teachers, including automating grading, providing analytics, and curriculum alignment. 
    \item\textbf{AI alignmrnt} The alignment of AI functions with curricula and standards by ensuring related to learning objectives, minimizing unnecessary or irrelevant content that can reduce extraneous load. Additionally, curriculum-aligned AI tools present information in a structured manner, following students’ progression from simpler to more complex concepts. This structuring helps manage the intrinsic load by ensuring that learners encounter material at an appropriate level of difficulty. This aspect specifically examines whether the AI function can address the standard when working to reduce extraneous load by presenting relevant content and managing intrinsic load by providing logically structured information. 
\end{itemize}

\section{Method}
\label{sec:method}
\begin{figure}
    \centering
    \includegraphics[width=1\linewidth]{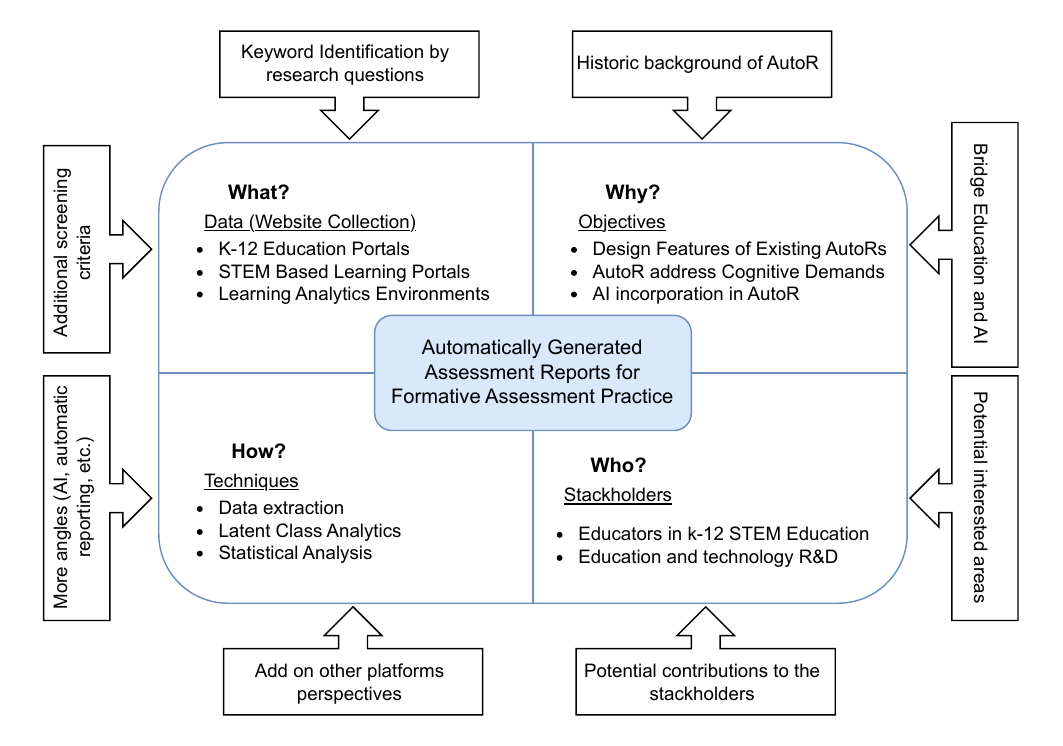}
    \caption{The analytics model of AutoR}
    \label{fig:analytics_model}
\end{figure}

Our review is built on \citet{borrego2014systematic} guidelines for data collection, keyword identification, study filtering, literature coding, and result synthesis. As AutoRs differ from academic articles, we adapted the method in the following ways: (a) We extended our search scope to multiple non-academic sources, including solicitation from teacher associations and official website documentation; (b) We further excluded AutoRs without practical usage or deployments; and (c) we intentionally framed our coding and result synthesis from the instructors’ perspective so that the synthesized results can guide the design of the AutoRs. The study follows an analytical approach (see Fig.~\ref{fig:analytics_model}), which involves data collection, objectives, techniques, and stakeholders. We collected K-12 AutoRs focusing on STEM education and include learning analytics following AutoR principles discussed earlier. The study aims to extract design features and methods to address cognitive demand and AI integration in AutoRs. We have used techniques such as data extraction from official sources of selected websites and Latent Class Analysis to classify AutoRs based on their features. 

\begin{figure}
    \centering
    \includegraphics[width=1\linewidth]{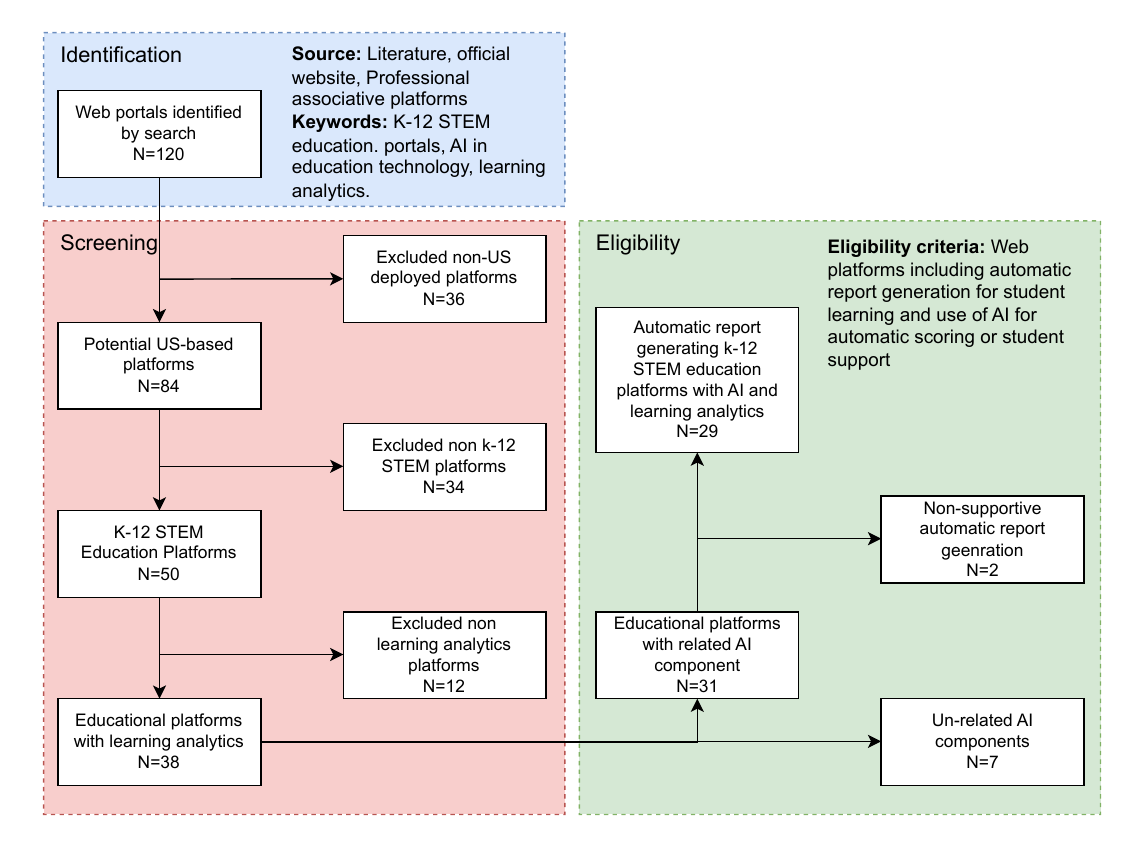}
    \caption{Platform selection process}
    \label{fig:selection_criteria}
\end{figure}

As seen in Figure 1, we collected 120 AutoRs from multiple sources (e.g., literature, official websites, and professional associations). We then reviewed the AutoRs using the following inclusion criteria: (a) applied to K-12 education and STEM subjects; (b) specifically targeted instructors as the main stakeholders; (c) designed for real-time authentic STEM classroom settings; (d) provided open access or fully-functional demos; (e) provided English GUI (graphic user interface) when other language GUIs available; (f) provided AI support in form of Chatbot or automatic scoring/assessment. After the initial screening, 38 AutoRs met the criteria, and then we excluded seven AutoRs due to their misalignment with the current study and two because of the nonexistence of supportive automatic report generation, leaving 29 AutoRs for further analyses. The exclusion and inclusion process overview can be seen in Fig.~\ref{fig:selection_criteria}.

\subsection{Data Extraction and Analysis}
Based on the conceptual framework of AutoRs design, we first developed the sub-dimensions for coding based on the design framework that we developed (Fig.~\ref{fig:design_framework}). The first two authors coded five randomly selected candidate AutoRs. Through discussion, we developed and revised the descriptions of the sub-dimensions (Tab.~\ref{tab:coding_rubric}). After finalizing the coding rubrics, we analyzed all AutoRs and extended a full review to the related publications to understand the usage and function in authentic learning environments. Two coders independently reviewed and coded all candidate AutoRs and related publications. We discussed discrepancies until we reached a consensus on all the codes.

\begin{table}[htp]
\centering
\rotatebox{90}{
\begin{tabular}{clp{4cm}p{8.5cm}}
\hline
\textbf{Dimension} & \textbf{Components} & \textbf{Codes} & \textbf{Explanation} \\
\hline
\multirow{20}{*}{\rotatebox[origin=c]{90}{Cognitive Demands}} & \multirow{15}{*}{Content} & Students' accomplishment on the tasks & The AutoR includes the record of the original student’s responses/inputs to the assessment items. \\
 &  & Students' timely performance & The AutoR captures the students’ endeavor spent on a certain assessment task and/or provides information on students' completion status. \\
 &  & Students' proficiency levels and/or distributions & The AutoR evaluates and presents the students' level of proficiency against the standards and/or rubrics defined by the AutoRs. \\
 &  & Learning performance expectations from educational standards & The AutoR provides official information aligning to the assessment items, for example, the disciplinary standards and learning expectations. \\
 &  & Learning progression & The AutoR provides explicit pathways of learning over periods of time, which include the lower anchor, higher anchor, and the pathways to make progress. \\
 &  & Historical information & The AutoR offers rich and detailed student records, including responses, learning practices, levels, etc., that show students’ learning over time. \\
 &  & Task difficulty & The AutoR provides diagnostic information on tasks and differentiates the tasks by levels. \\
\hline
\multirow{20}{*}{\rotatebox[origin=c]{90}{AI Embodiment}}& \multirow{20}{*}{ AI Functional Integration}& AI Embodiment Function & the AutoR integrated AI capabilities that enable the system to autonomously generate, analyze, and present data-driven insights while also interacting with users through an interface that simulates human-like decision-making and communication. \\
& & AI integration for reporting  & AutoR embedded AI components to automate the collection, analysis, and visualization of data, enabling the system to produce comprehensive and actionable reports with minimal human intervention. \\
& & AI function for Students  & The AI components in the AutoR offer real-time support to students in the classroom(e.g., personalized learning experiences, resources to individual learning needs ). \\
 & & AI function for Teachers  & The AI components in theAutoR offer real-time support to teachers (e.g., automating administrative tasks, offering personalized teaching resources, analyzing student performance data). \\
& & AI function alignment with curricula  & the AI component and resources are designed to integrate seamlessly with established educational standards and learning objectives. The resources content that is relevant, appropriate, and tailored to the specific learning outcomes set by the curriculum while also offering insights and assessments that align with educational goals \\
 \hline 
\end{tabular}
}

\caption{Coding Rubric}
\label{tab:coding_rubric_p1}
\end{table}

\begin{table}[htp]
\centering
\rotatebox{90}{
\begin{tabular}{clp{4cm}p{8.5cm}}
\hline
\textbf{Dimension} & \textbf{Components} & \textbf{Codes} & \textbf{Explanation} \\
\hline

 \multirow{23}{*}{\rotatebox[origin=c]{90}{Human Centered Design}} & \multirow{3}{*}{Information synthesis level} & Individual level & \\
 &  & Group level & \\
 &  & Class level & \\
 & \multirow{4}{*}{Depth of data mining} & Raw score & Students’ original responses and scoring. \\
 &  & Transformed score & An inference from the original responses and scoring, for example, the proficiency level. \\
 &  & Descriptive statistics & Common descriptive statistics include frequency, mean, median, sd, percentile ranks, quartile ranks, etc.\\
 &  & Inferential statistics & Common inferential statistics include t-test, confident interval, etc. \\
 & \multirow{3}{*}{User Functionality} & Live & The AutoR reports students’ learning process and task achievement in a timely fashion with a visualized format.  \\
 &  & Collaborative interaction & The AutoR serves as a platform that allows one or more of those interactions between (a) students and students, (b) teachers and students, (c) teachers and teachers, and (d) teachers and parents. \\
 &  & Filtering & The AutoR has the capability to individualize the AutoRs, allowing the instructor to personalize based on their own instructional concerns. \\
 & \multirow{3}{*}{Presentations} & Text & The AutoRs should present information in a series of sentences rather than keywords or phrases.  \\
 &  & Tables & \\
 &  & Plots & \\
 &  & Special Visual Aids \\
\hline
\end{tabular}
}

\caption{Coding Rubric (cont'd)}
\label{tab:coding_rubric_p2}
\end{table}

We first analyzed the observed frequency distribution of each characteristic in the 26 AutoRs using the R package, and we further classified these characteristics according to the distribution to answer research question 1. Then, we applied Latent Class Analysis (LCA) to analyze the latent patterns of the AutoRs to reveal the design features using the \texttt{poLCA} package \citep{linzer2011poLCA}in RStudio. Two dimensions – the Cognitive Demands and the Design Support were analyzed separately to answer research questions 2 and 3. For each dimension, we started with 1-class model and increased the number of classes by one to detect the most conceptually interpretative solution \citep{nylund2018ten}. Because of the small number of the reviewed AutoRs, the LCA method was applied only to give researchers possibilities of class distributions. Fit indices including AIC, BIC, and aBIC were reported but not applied as the main reasons for class determination. The final class is determined by researchers analyzing and comparing the differences and similarities in coding between dashbaords under different classess based on empirical understanding. 

\section{Results}
\label{sec:results}

\subsection{The Characteristics of Current AutoRs (frequency)}
Based on our review, the AutoRs reviewed balance the characteristics of the depth of data mining with user functionality to reflect their design purpose. From the design perspective, an AutoR might include one or more than one characteristic from each dimension. Even though an AutoR can include a few characteristics from the same dimension, it does not indicate this AutoR measured and presented overlapping information because each characteristic represents a different facet of the conceptual framework of AutoR design. However, they are intercorrelated to achieve the AutoR design purposes. Therefore, an AutoR can and should contain more than one characteristic from each dimension mentioned in Tab.~\ref{tab:dash_charcter}.

\begin{table}[htp]
\centering
\begin{tabular}{lp{5cm}cc}
\hline
Components & Characteristics &  Frequency & Percentage \\
\hline
\multirow{7}{*}{Cognitive demand content} & Student's accomplishment on the tasks & 29 & 100\% \\
 & Student's timely performance & 24 & 83\% \\
 & Student's proficiency level and/or distribution & 19 & 66\% \\
 & Learning performance expectation from standards & 14 & 48\% \\
 & Learning progression & 10 & 34\% \\
 & Historical information & 20 & 69\% \\
 & Task difficulty & 4 & 14\% \\
\hline
\multirow{3}{*}{AI Functional Integration} & AI embodiment function & 18 & 62\% \\
 & AI integration for reporting & 16 & 55\% \\
 & AI function for Students & 13 & 45\% \\
 & AI function for Teachers & 10 & 34\% \\
 & AI function alignment with curricula & 9 & 31\% \\
 \hline
\multirow{3}{*}{Information synthesizing level} & Individual level & 29 & 100\% \\
 & Group level & 8 & 28\% \\
 & Class level & 23 & 79\% \\
\hline
\multirow{4}{*}{Depth of data mining} & Raw score & 20 & 69\% \\
 & Transformed score & 12 & 41\% \\
 & Descriptive statistics & 24 & 83\% \\
 & Inferential statistics & 5 & 17\% \\
\hline
\multirow{3}{*}{User functionality} & Live & 18 & 62\% \\
 & Collaborative Interaction & 18 & 62\% \\
 & Filtering & 27 & 93\% \\
 & Use of Chatbot & 5 & 17\% \\
 & Use of Contextualized Chatbot & 3 & 10\%\\
\hline
\multirow{4}{*}{Presentation} & Text & 5 & 17\% \\
 & Table & 27 & 93\% \\
 & Plots & 20 & 69\% \\
 & Special visual-aids & 18 & 62\% \\
\hline
\end{tabular}
\caption{The AutoR characteristics distribution.}
\label{tab:dash_charcter}
\end{table}

Tab.~\ref{tab:dash_charcter} illustrates the frequency distribution of characteristics from the 26 AutoRs reviewed in this article. According to the coding scheme, we stratified the characteristics into three main categories:

\textbf{Elementary characteristics.} Observed in more than 70\% of AutoRs, indicating that these characteristics are default components of AutoR design. Examples include student accomplishments on tasks (100\%), individual level information synthesis (100\%), filtering (93\%), and data presentation in tables (93\%).

\textbf{Optional characteristics.} Observed in 40\%~70\% of AutoRs, these characteristics are selectively included to reflect unique features or specific design purposes. Examples include students' timely performance (83\%), class level information synthesis (79\%), historical information (69\%), and descriptive statistics (83\%).

\textbf{Advanced characteristics.} Observed in less than 40\% of AutoRs. These characteristics, although not widespread, serve specific design purposes, such as task difficulty (14\%) and learning progression (34\%), providing formative information about student learning. These features have the potential to synthesize bespoke information for formative assessment.

According to Tab.~\ref{tab:dash_charcter}, the design characteristics included in AutoR determine the affordance and magnitude of the students' learning information that a teacher can access.
First, the elementary characteristics of AutoR allow teachers to check the direct results of every student’s learning outcomes and the whole class's performance on tasks with a certain degree of flexibility. Two characteristics were found in all the samples: students’ accomplishments on the tasks and individual-level information synthesis. Additionally, most AutoRs provided functions such as filtering (93\%) and data presentation in tables (93\%). The majority of AutoRs that we reviewed capture students’ timely performance (83\%) while producing synthesized information at the class level (79\%). Lastly, most AutoRs (83\%) favor presenting results with descriptive statistics (e.g., mean, standard deviation, etc.).

Second, the optional characteristics of AutoR provide teachers with additional information beyond reporting direct results from classroom assessments. These design functions require teachers to interpret information about the progress of learning activities and students' cognitive learning performance. Historical information is included in 69\% of AutoRs, AI embodiment function in 62\%, and special visual aids in 62\%. The advanced characteristics, though less common, include features such as transformed scores (41\%) and inferential statistics (17\%), which offer deeper insights into student performance and learning progression.

Overall, the distribution of characteristics across the reviewed AutoRs shows a balanced integration of essential, optional, and advanced features, supporting diverse educational needs and facilitating comprehensive formative assessment.

Last, the advanced characteristics pre-synthesize information for teachers' interpretation. Grouping, task difficulty, inferential statistics, and text.

\subsection{The Cognitive Demands of Current AutoRs (LCA)}

\begin{table}[ht]
\centering
\caption{Latent class comparison model fit.}
\begin{tabular}{lccc}
\hline
Model & AIC & BIC & aBIC \\
\hline
1-Class & 839.8154 & 872.6305 & 803.3871 \\
2-Class* & 826.4430 & 893.4405 & 621.5960 \\
3-Class* & 812.3727 & 913.5526 & 305.7494 \\

\hline
\multicolumn{4}{l}{AIC is based on Akaike Information Criterion;}\\
\multicolumn{4}{l}{BIC is based on Bayesian Information Criterion;}\\
\multicolumn{4}{l}{aBIC is an adjusted BIC.}
\end{tabular}
\label{tab:model_fit}
\end{table}

The model fit statistics for the latent class analysis are presented in Table~\ref{tab:model_fit}. The table compares four models (1-Class to 4-Class) based on Akaike Information Criterion (AIC), Bayesian Information Criterion (BIC), and adjusted BIC (aBIC). The analysis aims to identify the optimal number of classes representing varying levels of cognitive demand across the components specified in Table~\ref{tab:coding_rubric_p1}. Here, Class 1 to Class 4 corresponds to increasing levels of cognitive demand.

The 1-Class model shows the highest AIC, BIC, and aBIC values, indicating a poor fit compared to models with multiple classes. The 2-Class and 3-Class models display lower AIC, BIC, and aBIC values, suggesting an improved fit with the data. The 3-Class model achieves the lowest values for all three criteria, particularly with an aBIC of 305.7494, supporting it as the best-fitting model. Although the 4-Class model was considered, its fit could not be adequately evaluated due to negative or non-applicable aBIC values, suggesting limitations in its validity.

These results suggest that the 3-Class model provides the best representation of the data, effectively capturing the variation in cognitive demand levels from low (Class 1) to high (Class 3) across the given components.

\begin{figure}[htp]
    \centering
    \includegraphics[width=1\linewidth]{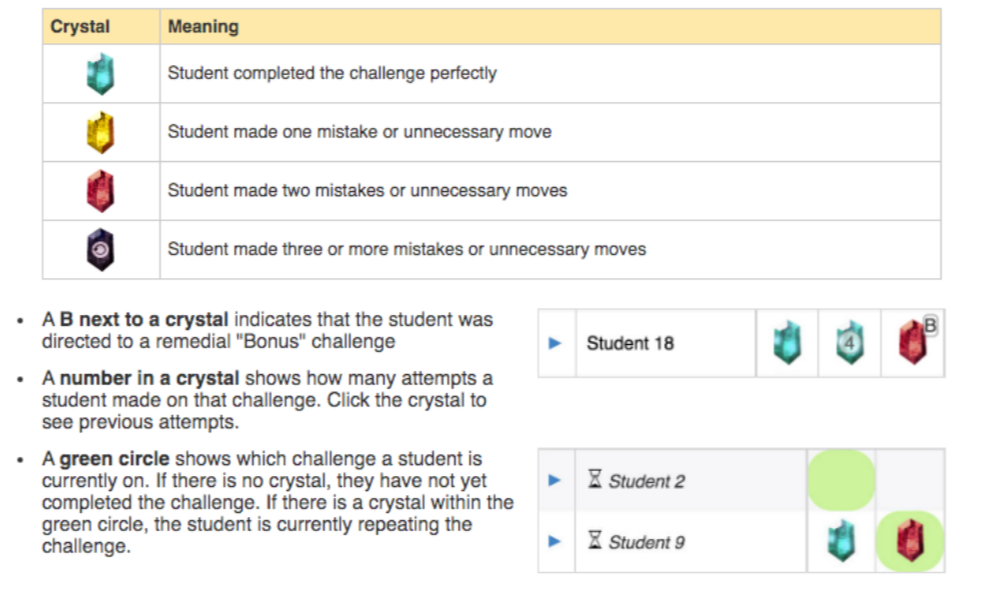}
    \caption{An example of Human centered AutoR demonstrating proficiency scores as Crystals.}
    \label{fig:autoR}
\end{figure}
\begin{figure}[htp]
    \centering
    \includegraphics[width=1\linewidth]{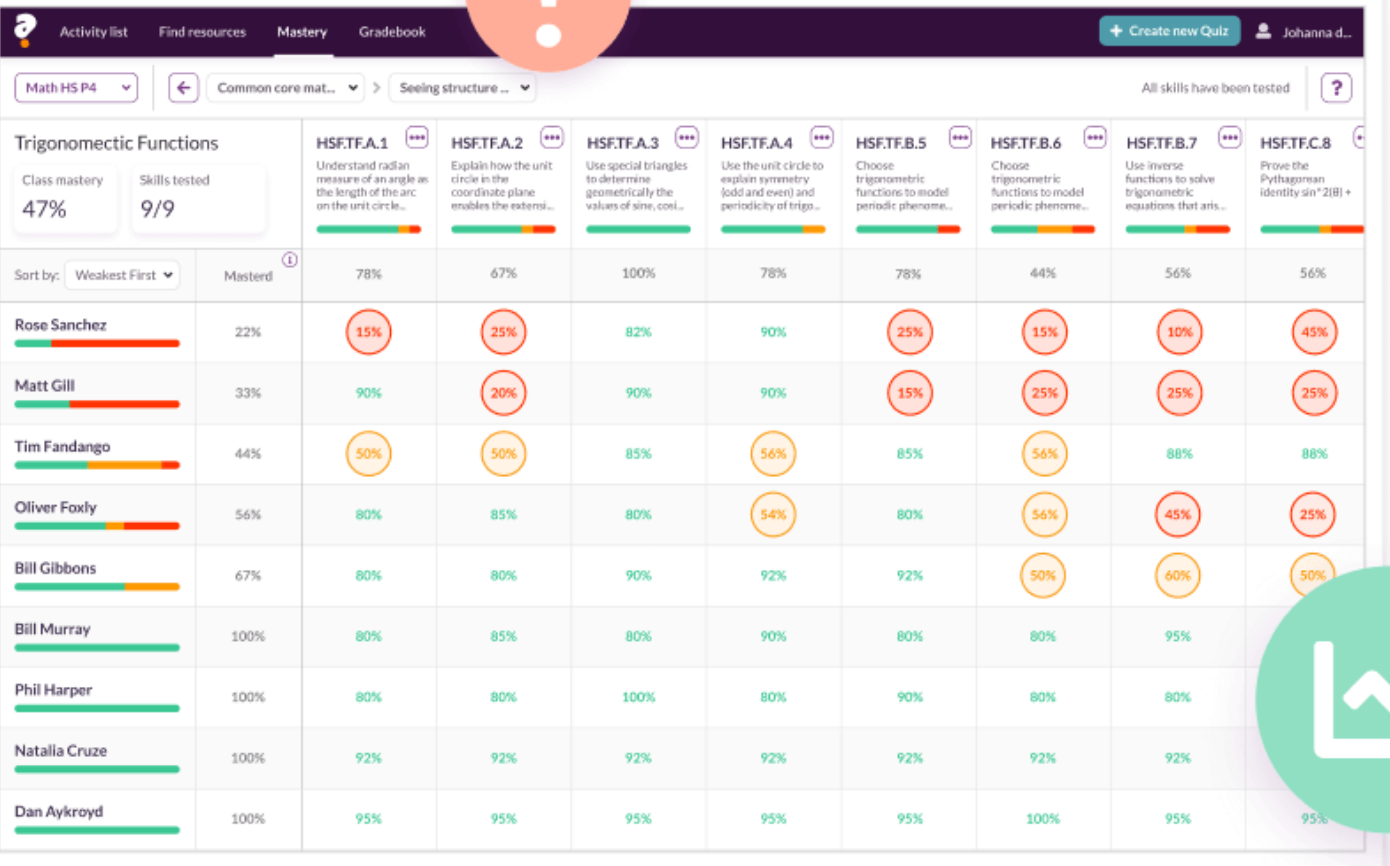}
    \caption{Quizalize is another AutoR providing comprehensive overview of student reports.}
    \label{fig:quizalize}
\end{figure}

Take Geniventure as a class 1 example. From the students’ progress report (Fig.~\ref{fig:autoR}), we observed this AutoR uses colored crystals to represent the level of student’s proficiency instead of in a lexical format, which requires teachers to understand the explanations of the crystals in advance. Similarly, another AutoR in the high cognitive load class - Quizalize (Fig.~\ref{fig:quizalize}) uses learning performance expectations from standards as the table crosshead. As the table provides diverse information about students’ achievement, expectation alignments, and historical information, it requires teachers to synthesize the data together to make interpretations, which means a high cognitive commands for teachers.

\subsection{The AI Integration(LCA)}

\begin{figure}
    \centering
    \includegraphics[width=1\linewidth]{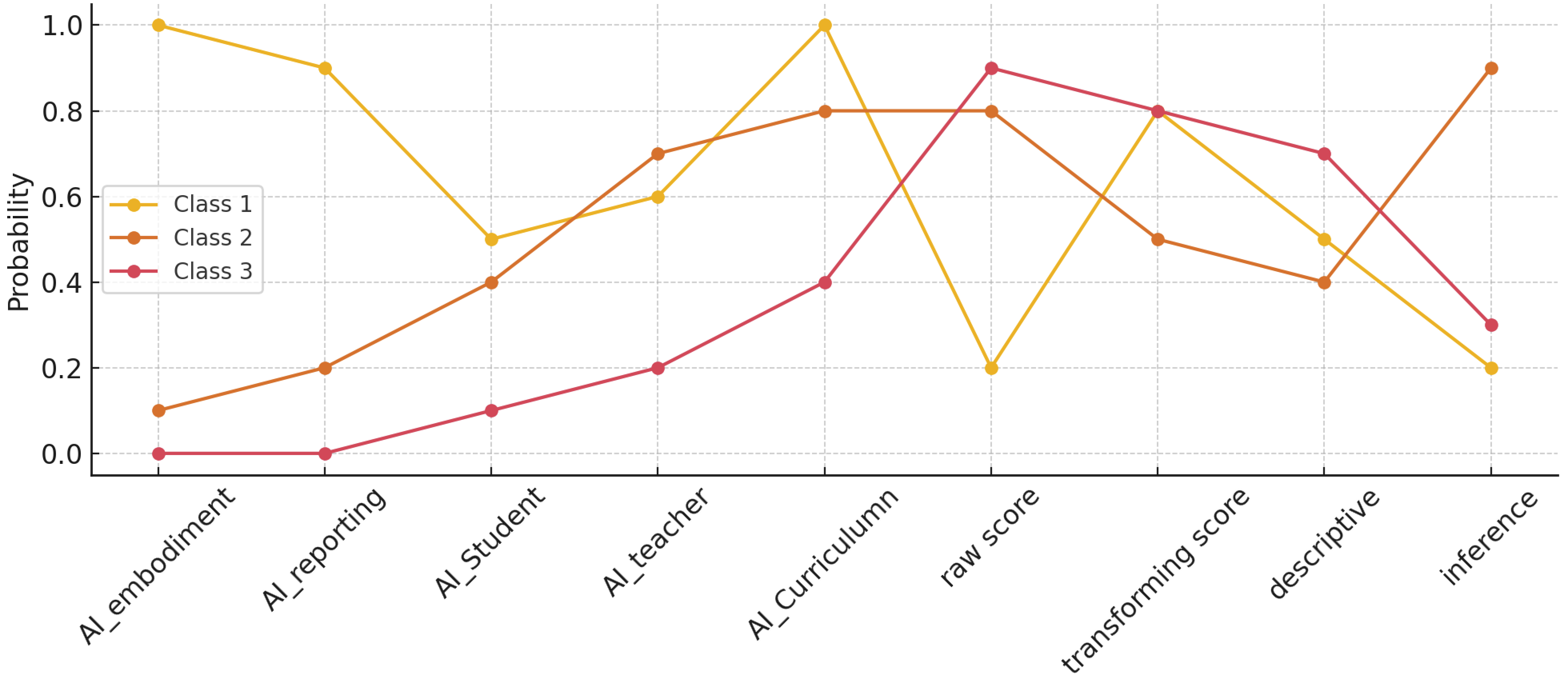}
    \caption{Plot of the conditional probability distribution of 3 classes in AI intergration}
    \label{fig:prob_dist_c2}
\end{figure}

The Latent Class Analysis (LCA) results for AI components are illustrated in Fig.~\ref{fig:prob_dist_c2}. The figure depicts the conditional probabilities for three distinct classes (Class 1, Class 2, and Class 3) across various AI components, including AI\_embodiment, AI\_reporting, AI\_Student, AI\_teacher, AI\_Curriculum, raw score, transforming score, descriptive, and inference.

Class 1 exhibits high probabilities for the initial components, such as AI\_embodiment and AI\_reporting, peaking at a probability of 1.0 for AI\_embodiment. However, its probabilities decline sharply for components like AI\_teacher and descriptive, reaching the lowest values for the inference component. The probabilities for Class 2 follow a distinct trajectory, starting low for AI\_embodiment but gradually increasing to high values for later components, such as transforming score and inference. This suggests that Class 2 aligns more closely with components focused on analytical and inferential capabilities. Class 3 shows a consistent increase in probabilities across the components, starting from AI\_embodiment and peaking at AI\_Curriculum. After this peak, the probabilities decline gradually, demonstrating a preference for earlier components but lower alignment with descriptive and inference aspects.

These findings highlight the differentiated alignment of the three classes with specific AI-related components, suggesting distinct latent structures in their integration and interpretation. The variability in class-specific probabilities underscores the nuanced roles of these components in the latent classification process.

\subsection{The Human Centered Design (LCA))}
\begin{table}[ht]
\centering
\caption{Latent class model fit statistics}
\begin{tabular}{lccc}
\hline
Model & AIC & BIC & aBIC \\
\hline
1-Class & 657.375 & 589.334 & 589.334 \\
2-Class & 609.324 & 549.232 & 522.294 \\
3-Class & 540.737 & 501.752 & 447.875 \\
4-Class & 550.209 & 512.315 & 431.500 \\
\hline
\multicolumn{4}{l}{AIC is based on Akaike Information Criterion;}\\
\multicolumn{4}{l}{BIC is based on Bayesian Information Criterion;}\\
\multicolumn{4}{l}{aBIC is an adjusted BIC.}
\end{tabular}
\label{tab:latent_class_fit}
\end{table}

Three classes were identified in Tab.~\ref{tab:latent_class_fit} from the LCA, including a class with full functionality (class 1), information highlight (class 2), and live feedback (class 3). Overall, all the three classes have similar probabilities on the use of filtering function and text and table presentation. The differences between the three types of AutoRs are mainly reflected in how they consider live and collaborate interaction functions, and how to use plot and visual-aids to present information.

\begin{figure}
    \centering
    \includegraphics[width=1\linewidth]{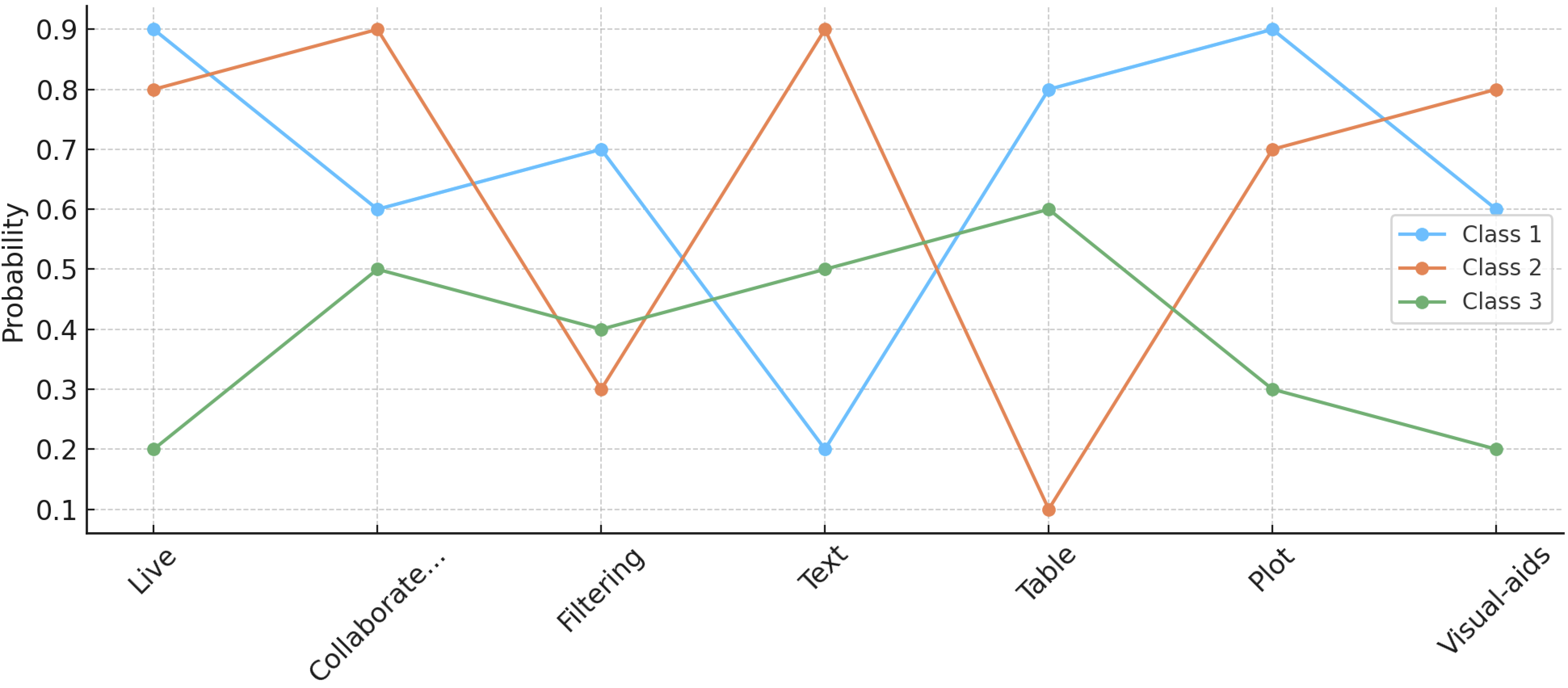}
    \caption{Plot of the probability distribution of 3 classes in design support.}
    \label{fig:prob_dist_c3}
\end{figure}

\begin{figure}[htp]
    \centering
    \includegraphics[width=1\linewidth]{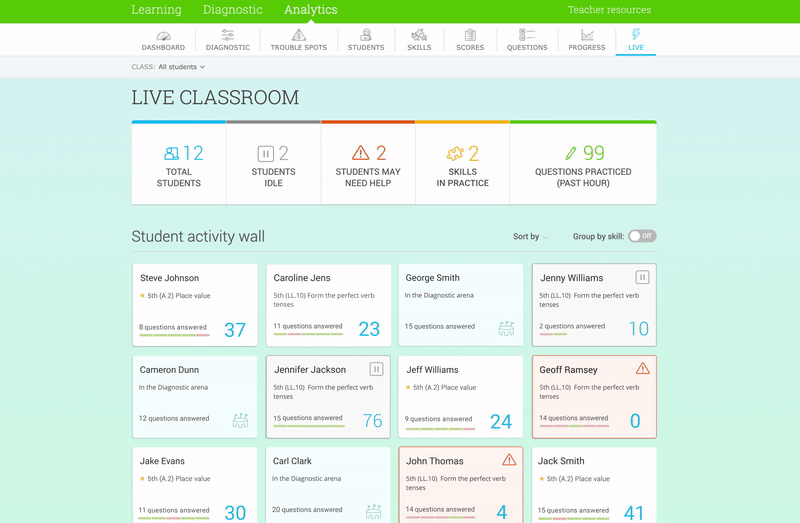}
    \caption{IXL is another AutoR delineating classroom level student reports.}
    \label{fig:ixl}
\end{figure}

Class 1 features high probabilities in all three characteristics (live, collaborative interaction, and filtering) specified in the User Functionality dimension. Therefore, we categorized AutoRs in Class 1 as a fully functional group. Fig.~\ref{fig:ixl} shows the live classroom report function from IXL. The report provides teachers with dynamic information about students' task practice. The report also highlights students who encountered difficulties in time, therefore, teachers can provide additional assistance accordingly. At the same time, this AutoR handles teacher-student messages with the interactive function, which allows students to receive individual feedback in real-time. For the information presentation function, Fig.~\ref{fig:prob_dist_c3} indicates that Class 1 AutoRs has a relatively low probability of using visual aids (48.6\%) except for plots and tables to present the information. Such a relatively low probability is reasonable as various user functions already require teachers’ attention. In sum, the design of the fully functional group supports teachers to manage and monitor students’ learning activities in class.

\begin{figure}
    \centering
    \includegraphics[width=1\linewidth]{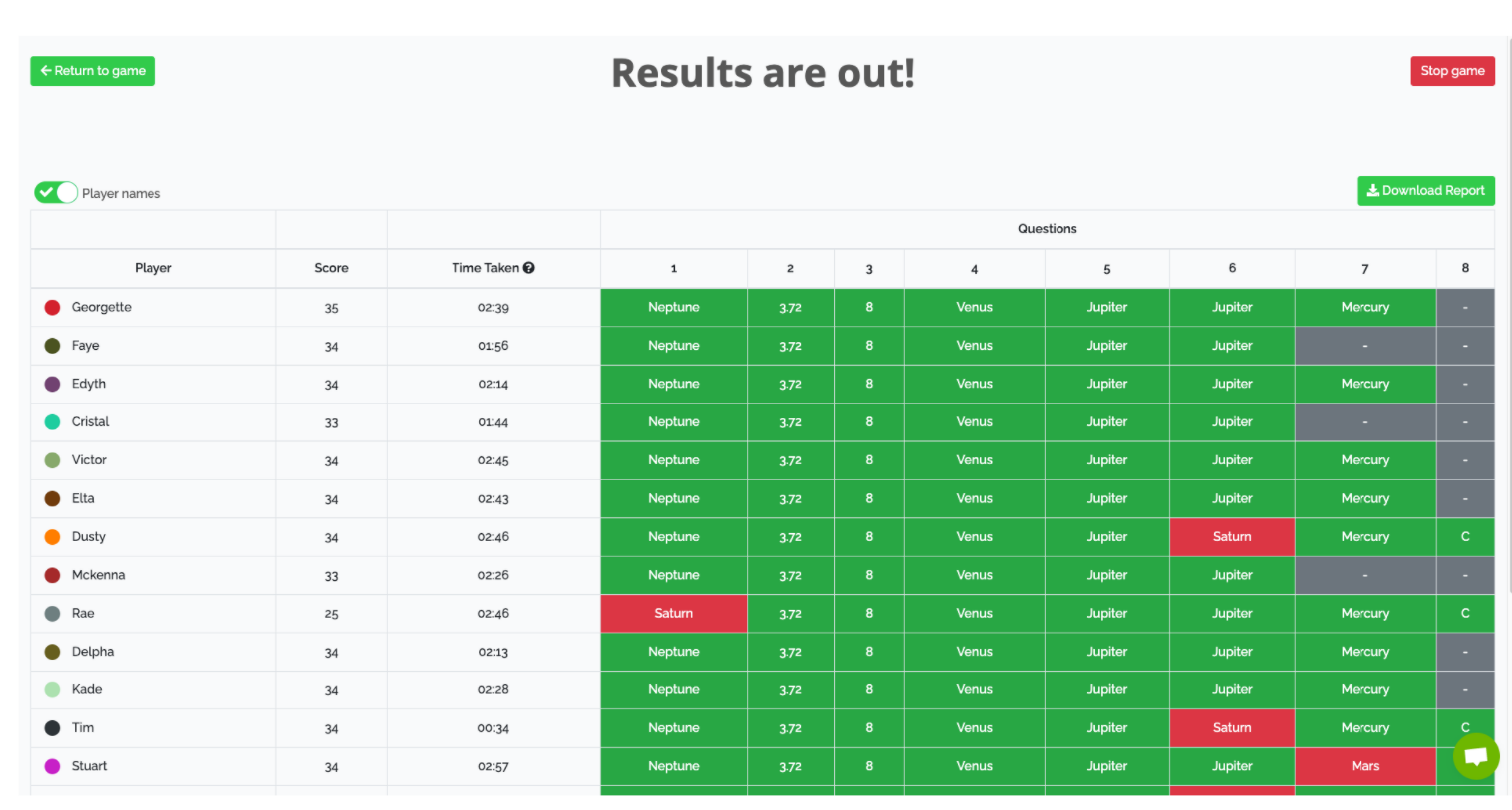}
    \caption{Quiz Whizzer shows individual student's performance for specific tasks along with the time taken for each student.}
    \label{fig:quizz_whizzer}
\end{figure}

The design of AutoRs in class 2 supports teachers by providing data information. Compared with Class 1 AutoRs, class 2 AutoRs demonstrated a higher probability of using visual aids to present information to teachers in real-time. For example, Quizzwhizzer (Fig.~\ref{fig:quizz_whizzer}) presents results from a classroom game activity with contrasting colors (red and green). The visual-aid highlights the wrong answer in red from the table, which helps teachers easily eyeball the result in time. However, this AutoR does not stress live and interactive functions for teachers to communicate with students. Class 2 AutoRs have a low probability of providing live (16.2\%) and interaction (28.4\%) functions. Further, we observed a high probability of filtering function (91.5\%). Both the filtering function and visual-aid presentation support teachers in organizing and synthesizing the data according to their needs. Therefore, class 2 AutoRs are more task results-oriented than the fully functional group. 

\begin{figure}
    \centering
    \includegraphics[width=1\linewidth]{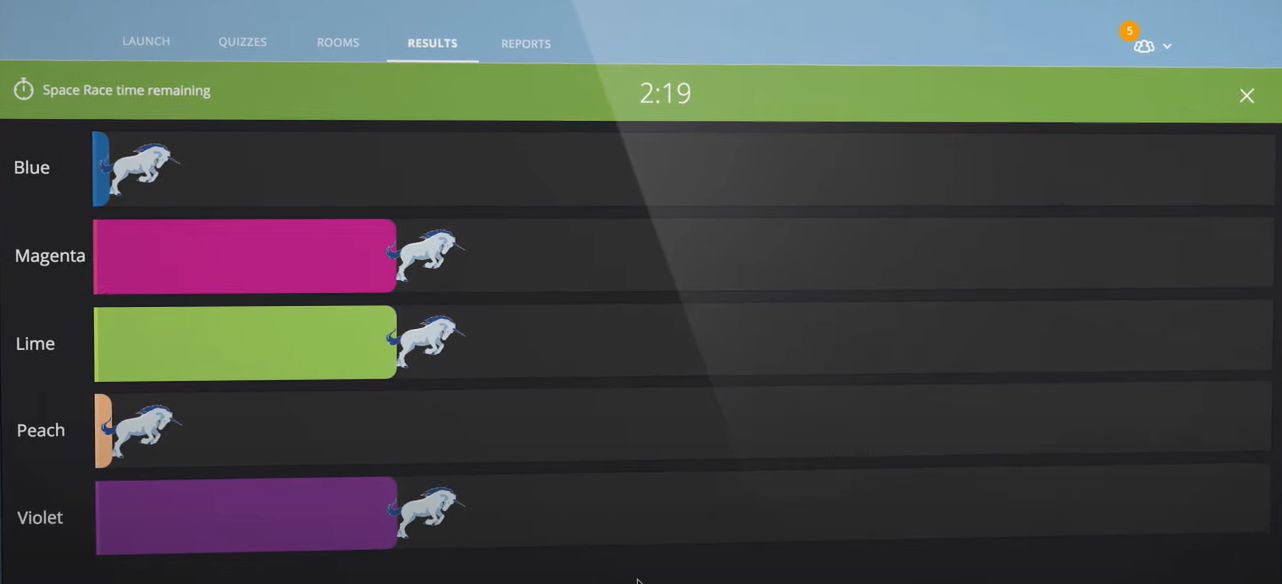}
    \caption{Socrative is an interactive AutoR showing comparative results in the form of growing histogram for time elapes of each student.}
    \label{fig:socrative}
\end{figure}

Class 3 AutoRs provide simultaneous feedback to teachers and students on their ongoing task activities, making AutoRs an integral part of the activity rather than a tool for the activity summary. AutoRs in class 3 stressed the live function (100\%) more than the interaction function (52.2\%). In addition to teachers’ monitoring, the live function is used in providing real-time feedback to students during a class activity. For example, Fig.~\ref{fig:socrative} depicts live and interaction functions from Socrative. The panel provides real-time team task progress, which is available to both students and the teacher. Even though the class 3 AutoRs aim to provide simultaneous feedback on task activities, it is worth noticing that the AutoR from this class did not feature plots or special visual aids. 

\section{Discussion}
\label{sec:disccusion}

This study reviewed the use of Automated Reporting Systems (AutoRs) in STEM classrooms to support formative assessment practices. Guided by cognitive load theory, we developed a design framework to evaluate the feasibility and usability of AutoRs’ characteristics. Our analysis revealed that AutoRs possess shared and distinct features, which can be categorized into three tiers: elementary, optional, and advanced features. These shared and unique characteristics are explained by the cognitive demands and functional design requirements associated with each AutoR. This paper discusses the classification of these features from both cognitive and functional design perspectives.

The cognitive demands of AutoRs determine the baseline cognitive load required for practitioners to effectively use them in classrooms. Higher cognitive load activities demand practitioners invest additional time and effort to process the provided data, particularly when AutoRs present rich and multifaceted content \citep{kirschner2009cognitive}. Our analysis categorized AutoRs into two groups: high cognitive load and low cognitive load. High cognitive load AutoRs offer a wide array of processed information, including expected student performance, learning progressions, and historical data. These features are considered valuable for teachers to accurately evaluate student learning \citep{chen2016robust}. However, such features also require deep cognitive engagement, as teachers must perform additional steps to interpret the results.

Specifically, high cognitive load AutoRs place demands on teachers in three key areas. First, interpreting information such as student proficiency levels and expected learning performance necessitates teachers applying their pedagogical content knowledge (PCK). Second, transforming scores and group information requires teachers to retrieve knowledge from their long-term memory, such as statistical concepts, grouping strategies, and pedagogical considerations. Third, this retrieval process can increase cognitive load when managing and synthesizing information in working memory \citep{sweller2016working}. Despite these challenges, some cognitive demands are unavoidable due to the complexity and significance of the information being presented.

Conversely, low cognitive load AutoRs reduce cognitive demands on teachers but often lack critical features. This trade-off explains why certain characteristics of AutoRs are categorized as optional. While some features are deemed elementary to AutoR design, designers must carefully balance their inclusion against the overall usability of the system.

To mitigate cognitive load, two design strategies can be employed. First, incorporating elementary design functions, such as filtering and tabular presentation formats, allows for preliminary information screening and intuitive synthesis. However, these approaches have limitations. For instance, while filtering and tabular displays reduce cognitive load by aggregating information, teachers must still synthesize and process the presented data.

Second, to enhance information synthesis, more advanced presentation formats, such as visual aids, plots, and textual descriptions, can be utilized. From a theoretical perspective, presenting classroom assessment information in textual format has been shown to reduce cognitive load, particularly when the information is dense and interrelated \citep{clark2011efficiency, destefano2007cognitive}. Despite these advantages, few AutoRs currently adopt text-based formats. Similarly, plots and visual aids are used sparingly, highlighting significant potential for leveraging design functions to reduce teachers’ cognitive load.

While design functions aim to alleviate the cognitive burden associated with AutoRs, they can simultaneously expand the usability goals of these systems. However, this expansion often introduces additional cognitive demands on teachers. For example, in the highest functionality group (Class 3), AutoRs provide live and collaborative interaction capabilities. These features assist teachers in organizing and monitoring classroom activities while offering real-time feedback to individual students. While timely feedback has been shown to positively influence student learning outcomes \citep{chen2017knowledge}, using such interactive features requires teachers to possess pre-existing operational knowledge. Additionally, Class 3 includes the ability to share feedback between teachers and students, which further requires teachers to process student performance data while considering the implications of sharing feedback simultaneously.

This highlights another critical design trade-off for AutoRs: the extent to which these systems should address non-academic dimensions of student performance, such as behavior and emotional states. The inclusion of such features introduces further complexity to the design and use of AutoRs, necessitating careful consideration of their role in supporting classroom assessment practices.

\section{Conclusion}
This study applied a design framework grounded in cognitive load theory to evaluate the theoretical cognitive demands of current Automated Reporting Systems (AutoRs) in educational settings. The findings highlight that while AutoRs provide valuable formative assessment information, they often impose a high initial cognitive load on teachers, particularly during the interpretation and application of the data. To mitigate this cognitive burden, the study suggests that AutoRs should adopt diverse and optimized information presentation formats, such as text, plots, and visual aids. However, our review indicates that these approaches remain underutilized in most of the AutoRs analyzed. Enhancing these presentation formats could significantly alleviate teachers' cognitive load, making the systems more accessible and efficient. Additionally, the incorporation of advanced functionalities, such as live and interactive features, could extend the utility of AutoRs to a broader range of teaching and learning scenarios. However, while these features enhance the systems’ capabilities, they introduce new operational cognitive demands for teachers, necessitating thoughtful consideration during the design process to balance usability and functionality. This study underscores the need for a more comprehensive integration of cognitive load principles in the design of AutoRs to better support teachers in formative assessment practices.

\section{Acknowledgement}
This presentation was funded by the National Science Foundation(NSF) (Award no. 2101104, 2138854). 
Any opinions, findings, conclusions, or recommendations expressed in this material are those of the authors and do not necessarily reflect the views of the NSF.

\section*{Declaration of generative AI and AI-assisted technologies in the writing process}

During the preparation of this work the authors used ChatGPT in order to check grammar and polish the wordings. After using this tool/service, the authors reviewed and edited the content as needed and take full responsibility for the content of the publication.

\bibliography{sn-bibliography}

\end{document}